# An Updated Lagrangian Particle Hydrodynamics (ULPH) Implementation of Heat Conduction Model for Weakly-Compressive Fluid

Junsong Xiong[a,b], Zhen Wang[a,b], Xin Lai[a,b,*], Lisheng Liu[a], Xiang Liu[b]
a, Hubei Key Laboratory of Theory and Application of Advanced Materials Mechanics, Wuhan University of Technology, Wuhan 430070, China
b, Department of Engineering Structure and Mechanics, Wuhan University of Technology, Wuhan 430070, China
* Corresponding author: laixin@whut.edu.cn

## Abstract

Heat conduction is quite common in natural, industrial, and military applications. In this work, the updated Lagrangian particle hydrodynamics (ULPH) theory, is utilized and applied to solve heat conduction problems. Since heat conduction is a second-order problem, the high-order ULPH theory is employed to establish the governing equations of heat conduction in ULPH, which is then validated using various numerical simulations. In this work, numerical simulations have been carried out to solve both static heat conduction problems and dynamic heat convection problems. The results show good accuracy and capability of the ULPH heat conduction model, suggesting promising prospects of the ULPH theory in multiphysics problems. The findings of this paper suggest that ULPH is effective in addressing convective heat transfer problems.

## 1. Introduction

Heat transfer phenomena are pervasive across diverse domains, including aerospace, automotive industry, electronic engineering, and so on[1][2][3][4]. Notable applications encompass heat exchangers[5], nuclear reactors[6], chip cooling[7], among others. Acquiring precise thermal flow field data through experimental methods proves challenging when investigating heat transfer. Consequently, numerical simulation emerges as a potent complementary tool, facilitating the comprehension of the flow field evolution at any given instance. Traditional numerical research methods, such as Finite Volume Method (FVM)[8], Finite element method (FEM)[9][10], and more, have been widely applied in the field of heat transfer. However, their applicability becomes constrained in the face of intricate challenges like complex flow fields, multi-physics coupling, and highly non-linear phenomena[11]. In light of these limitations associated with grid-based methods, there arises a compelling necessity to explore innovative numerical approaches for more effectively tackling these intricate issues.

Lately, meshless methods have emerged as a focal point within computational fluid dynamics. Their paramount advantage resides in their capacity to perform fluid simulations devoid of a predefined grid, ensuring enhanced adaptability and flexibility. Meshless methods prove exceptionally adept in managing intricate, unstructured fluid computations, offering robust assistance for quantitative analysis of complex fluid phenomena. Smoothed Particle Hydrodynamics (SPH) is indeed a notable and early meshless method, originally proposed by Lucy[12] in 1977, initially applied in astrophysics. As a meshless particle method, SPH discretizes the entire medium of the flow field into a system of particles. These particles embody all mechanical properties and move in an arbitrary manner guided by fluid dynamics principles. Fundamentally, SPH relies on the theory of kernel function approximation for calculating spatial derivatives [13]. This enables the computation of gradients, divergences, and other physical field properties using integral forms. However, one limitation of SPH emerges from the prevalent usage of discontinuous or non-smoothly differentiable kernel functions, leading the prediction of second derivatives less accurate compared to the precision achieved with first-order derivatives.

Cleary and Monaghan[14] made an alteration to the standard SPH formulation and formulated a set of rules for constructing isothermal boundaries, resulting in an accurate SPH conduction solution. Suprijadi et al.[15] used the SPH method to simulate the melting process of ice at different ambient temperatures. Hosain et al.[16][17] used SPH to perform convective heat transfer simulations of laminar flow, verified the feasibility of the method, and implemented the heat conduction equation in the open-source code DualSPHysics based on SPH, which was applied to different research cases to demonstrate the potential of the gridless method in industrial applications involving heat transfer.

Inspired by SPH and Peridynamics (PD), a novel meshless method (ULPH) has demonstrated promising results in various applications, including multiphase flow, surface flow, and solid-water interaction[18][19][20][21]. The development of a higher-order ULPH theory by Yan et al.[22] addresses the need for arbitrary-order derivatives. Consequently, theoretically, ULPH is anticipated to deliver higher accuracy and improved stability compared to traditional SPH methods, without necessitating additional restrictions or manipulation of thermal boundaries.

Up to this point, ULPH has not yet been extended to the field of heat transfer. The present work aims to bridge this gap by deriving the ULPH heat transfer equations and establishing a model framework for ULPH heat transfer. This endeavor seeks to validate the feasibility and advantages of applying ULPH theory in the context of heat transfer.

This paper is arranged as follows: In section 2, we shall introduce the fundamental theory and governing equations of ULPH. And the heat transfer equation for ULPH is derived. This section also encompasses an exploration of temporal integration schemes and boundary conditions. Numerical validation is conducted through comparisons with various numerical methods in section 3, including ULPH and Finite Volume Method (FVM). Several numerical instances are provided to corroborate the stability and accuracy of the ULPH model. Section 4 encapsulates a comprehensive summary of the computational research on ULPH applied to heat transfer problems, accompanied by a discussion on potential avenues for further development.

## 2. Updated Lagrangian particle hydrodynamics (ULPH) method

In this section, we provide a synopsis of the ULPH formulations for weakly compressible fluids and advance the ULPH heat transfer differential equations. Given that heat transfer constitutes a prototypical second-order problem, the application of second-order ULPH theory becomes imperative, enabling the direct computation of the Laplacian operator for temperature without the necessity of variational formulations like SPH[14].

### 2.1 Governing equations

The model investigated in this paper is based on the Weakly Compressible Smoothed Particle Hydrodynamics (WCSPH) model proposed by Monaghan[23]. To satisfy the weakly incompressible condition, it is assumed that

$$Ma = \frac{\|v\|}{c_0} \leq 0.1 \tag{1}$$

Here, $Ma$ represents the Mach number of fluid, $v$ denotes the fluid velocity, and $c_0$ is the artificial sound speed. According to Eq. (1), the artificial sound speed must satisfy:

$$c_0 \geq 10 \max(\|v\|) \tag{2}$$

The general form of the Lagrangian formulation of weakly compressible Newtonian fluid control equations is as follows:

continuity equation:

$$\frac{D\rho}{Dt} = -\rho \nabla \cdot v \tag{3}$$

momentum equation:

$$\frac{Dv}{Dt} = \frac{1}{\rho} \nabla \cdot \sigma + f \tag{4}$$

where $\sigma$ is the Cauchy stress tensor, with the expression: $\sigma = -pI + \tau$, which is the summation of a pressure term $-pI$ (hydrostatic stress or volumetric stress) and a viscosity term $\tau$ (deviatoric stress). $f$ denotes the volume force.

The pressure field ($p$) is calculated by using the equation of state based on the instantaneous particle density ($\rho$) and the initial density ($\rho_0$) of the fluid[23]:

$$p = B\left[\left(\frac{\rho}{\rho_0}\right)^\gamma - 1\right] \tag{5}$$

where $\gamma$ is the exponential coefficient, which is 1.4 for gas and 7 for water; and B is related to the sound speed and mass density by $B = \rho_0 c_0^2 / \gamma$.

In order to simplify the model, we just consider the heat conduction and heat convection processes in energy equation. The thermal radiation process is not considered in this study. The following heat energy equation is used:

$$c_p \frac{DT}{Dt} = \frac{1}{\rho} \nabla \cdot (\kappa \nabla T) \tag{6}$$

where $c_p$ is specific heat, $T$ is temperature and $\kappa$ is the thermal conductivity. When the thermal conductivity is constant, Eq. (6) can be deformed as:

$$c_p \frac{DT}{Dt} = \frac{\kappa}{\rho} \nabla^2 (T) \tag{7}$$

It is worth mentioning that ULPH is a Lagrangian method. Thus, in the ULPH

simulation, thermal convection is naturally included in the random derivative to the left of Eq. (6).

## 2.2 ULPH scheme

In ULPH method, nonlocal differential operators are used to calculate divergence, gradient, and curl[18]. These nonlocal differential operators[18] are respectively defined as follows:

$$\nabla_I \cdot (\bullet) := \int_{\mathcal{H}_I} \omega(\boldsymbol{x}_{IJ}) \Delta(\bullet) \cdot M_I^{-1} \boldsymbol{x}_{IJ} dV_J \tag{8}$$

$$\nabla_I \otimes (\bullet) := \int_{\mathcal{H}_I} \omega(\boldsymbol{x}_{IJ}) \Delta(\bullet) \otimes M_I^{-1} \boldsymbol{x}_{IJ} dV_J \tag{9}$$

$$\nabla_I \times (\bullet) := \int_{\mathcal{H}_I} \omega(\boldsymbol{x}_{IJ}) M_I^{-1} \boldsymbol{x}_{IJ} \times \Delta(\bullet) dV_J \tag{10}$$

where ( $\bullet$ ) denotes an arbitrary field function, $\Delta(\bullet) := (\bullet)_J - (\bullet)_I$. The operators ( $\nabla \cdot$ )、( $\nabla \otimes$ )、( $\nabla \times$ )represent the nonlocal divergence, gradient and curl operators, respectively. $\mathcal{H}_I$ represents the support domain of particle $I$ with a circular domain in two-dimensional space or a spheroidal domain in three-dimensional space. particle $J$ denotes the family member of particle $I$, i.e $\boldsymbol{x}_J \in \mathcal{H}_I$. $\omega(\boldsymbol{x}_{IJ})$ [22] represents the kernel function, and $\boldsymbol{x}_{IJ} = \boldsymbol{x}_J - \boldsymbol{x}_I$ represents the relative position vector of the particle. $M_I$ is the moment matrix with a symmetrical property defined as

$$M_I := \int_{\mathcal{H}_I} \omega(\boldsymbol{x}_{IJ}) \boldsymbol{x}_{IJ} \otimes \boldsymbol{x}_{IJ} dV_J \tag{11}$$

As a meshfree particle method, the ULPH method requires the discretization of the computational domain into particles with physical properties. Therefore, nonlocal differential operators presented in Eqs. (8) (9) and (10) can be expressed in the particle approximation form as follows:

$$\nabla_I \cdot (\bullet) := \sum_{J=1}^{N} \omega(x_{IJ})(\Delta(\bullet)) \cdot (M_I^{-1} \boldsymbol{x}_{IJ}) V_J \tag{12}$$

$$\nabla_I \otimes (\bullet) := \sum_{J=1}^{N} \omega(x_{IJ})(\Delta(\bullet)) \otimes M_I^{-1} \boldsymbol{x}_{IJ} V_J \tag{13}$$

$$\nabla_I \times (\bullet) := \sum_{J=1}^{N} \omega(x_{IJ})(M_I^{-1} \boldsymbol{x}_{IJ}) \times (\Delta(\bullet)) V_J \tag{14}$$

Using the above nonlocal differential operators and peridynamics theory[24], the governing equations can be rewritten in the ULPH discrete form[19] as following :

$$\frac{D\rho_I}{Dt} = -\rho_I \sum_{J=1}^{N} \omega(\boldsymbol{x}_{IJ})(v_J - v_I) \cdot M_I^{-1} \boldsymbol{x}_{IJ} V_J \tag{15}$$

$$\frac{Dv_I}{Dt} = \frac{1}{\rho_I} \sum_{J=1}^{N} \omega(\boldsymbol{x}_{IJ})(\sigma_J M_J^{-1} + \sigma_I M_I^{-1}) \boldsymbol{x}_{IJ} V_J + f_I \tag{16}$$

Since the second derivative of temperature needs to be calculated when calculating the heat transfer equation (in Eq. (7)), the higher-order ULPH theoretical formula[22] is required. Taking the two-dimensional second-order ULPH theoretical formula as an example, we reconstruct the heat transfer equation.

The formula for calculating the nonlocal differential operator $d(x)$ is as follows:

$$d(x_I) = M^{-1}(x_I) \sum_{J=1}^{N} \omega(x_J - x_I) q(x_J - x_I)(u(x_J) - u(x_I)) V_J \quad (17)$$

where, $q(x_J - x_I)$ is the basis of the polynomial, take the two-dimensional quadratic basis:

$$q(x_J - x_I) = [x_{1IJ}, x_{2IJ}, (x_{1IJ})^2, x_{1IJ} x_{2IJ}, (x_{2IJ})^2] \quad (18)$$

where, $x_{1IJ} = x_{1J} - x_{1I}$, $x_{2IJ} = x_{2J} - x_{2I}$.

$M(x_I)$ represents the moment matrix with symmetric properties defined as

$$M(x_I) = \sum_{J=1}^{N} \omega(x_J - x_I) q(x_J - x_I) q^T(x_J - x_I) V_J \quad (19)$$

Thus, the first and second derivatives of the function $u(x)$ based on the basis of a quadratic polynomial $q(x)$ are as follows:

$$d(x_I) = \begin{bmatrix} \dfrac{\partial u}{\partial x_1} \\ \dfrac{\partial u}{\partial x_2} \\ \dfrac{\partial^2 u}{\partial x_1^2} \\ \dfrac{\partial^2 u}{\partial x_1 \partial x_2} \\ \dfrac{\partial^2 u}{\partial x_2^2} \end{bmatrix}_{x=x_I} = \begin{bmatrix} (1\,0\,0\,0\,0) M^{-1}(x_I) \sum_{J=1}^{N} \omega(x_J - x_I) q(x_{IJ})(u(x_J) - u(x_I)) V_J \\ (0\,1\,0\,0\,0) M^{-1}(x_I) \sum_{J=1}^{N} \omega(x_J - x_I) q(x_{IJ})(u(x_J) - u(x_I)) V_J \\ (0\,0\,2\,0\,0) M^{-1}(x_I) \sum_{J=1}^{N} \omega(x_J - x_I) q(x_{IJ})(u(x_J) - u(x_I)) V_J \\ (0\,0\,0\,1\,0) M^{-1}(x_I) \sum_{J=1}^{N} \omega(x_J - x_I) q(x_{IJ})(u(x_J) - u(x_I)) V_J \\ (0\,0\,0\,0\,2) M^{-1}(x_I) \sum_{J=1}^{N} \omega(x_J - x_I) q(x_{IJ})(u(x_J) - u(x_I)) V_J \end{bmatrix} \quad (20)$$

Thus, the two-dimensional Laplace operator with respect to temperature T is obtained:

$$\nabla^2 T = \frac{\partial^2 T}{\partial x_1^2} + \frac{\partial^2 T}{\partial x_2^2} = (0\,0\,2\,0\,0) M^{-1}(x_I) \sum_{J=1}^{N} \omega(x_J - x_I) q(x_{IJ}) T_{IJ} V_J \quad (21)$$

$$+ (0\,0\,0\,0\,2) M^{-1}(x_I) \sum_{J=1}^{N} \omega(x_J - x_I) q(x_{IJ}) T_{IJ} V_J$$

Substituting into Eq. (7), the ULPH heat transfer equation is obtained:

$$\frac{DT}{Dt} = \frac{\kappa}{\rho \cdot c_p} \nabla^2 T = \frac{\kappa}{\rho \cdot c_p} \begin{bmatrix} (0\,0\,2\,0\,0) M^{-1}(x_I) \sum_{J=1}^{N} \omega(x_J - x_I) q(x_{IJ}) T_{IJ} V_J \\ + (0\,0\,0\,0\,2) M^{-1}(x_I) \sum_{J=1}^{N} \omega(x_J - x_I) q(x_{IJ}) T_{IJ} V_J \end{bmatrix} \quad (22)$$

## 2.3 Boundary condition

For solid-wall boundary, the application of virtual particles is used to deal with the boundary problem, by applying three layers of boundary particles to prevent the penetration of fluid particles. And make sure that the neighborhood near the boundary particles is intact, as shown in Fig. 1.

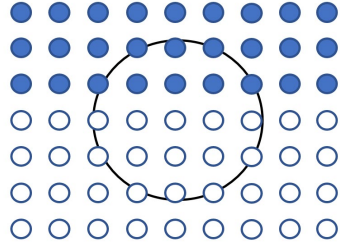

Fig. 1 Fluid particles ( ● ) and boundary particles ( ○ ).

The position of the boundary particles remains constant, while other physical information is derived from the fluid particles in their neighborhood. By imposing the no-slip boundary condition, the velocity[25] of the solid-walled virtual particles is:

$$v_w = 2v_i - \tilde{v}_i \tag{23}$$

where $v_i$ is the preset boundary velocity, and $\tilde{v}_i$ is the velocity that the solid-walled virtual particle interpolates from the fluid particle in the support domain, with its expression:

$$\tilde{v}_i = \frac{\sum_j^N v_j w_{ij}}{\sum_j^N w_{ij}} \tag{24}$$

here $i$ represents the boundary particle, and $j$ represents a fluid particle in the neighborhood of $i$.
Similarly, the pressure of solid-walled particles also varies from the interpolation to the following:

$$p_i = \frac{\sum_j^N p_f w_{ij} + (g - a_i) \cdot \sum_j^N \rho_j x_{ij} w_{ij}}{\sum_j^N w_{ij}} \tag{25}$$

where $g$ is the acceleration of gravity, $a_i$ is the prescribed acceleration of the solid boundary while $a_i = 0$ is used for the fixed solid boundary condition.

## 2.4 Time integration

For the selection of time step, the number of CFL conditions must be satisfied. Referring to[26], the selection expression for the time step $\Delta t$ is:

$$\Delta t \leq \min\left(\Delta t_f, \Delta t_v, \Delta t_c\right) \tag{26}$$

where

$$\begin{aligned}
\Delta t_f &= \alpha_f \min \sqrt{h_I / \|a_I\|}, \\
\Delta t_v &= \alpha_v \min\left(h_I^2 / v\right), \\
\Delta t_c &= \alpha_c \min\left(h_I / c_0\right).
\end{aligned} \tag{27}$$

here $\alpha_f$ =0.25, $\alpha_v$ =0.125, $\alpha_c$ =0.25. $a_I$ represents the acceleration of particle $I$, $v$ denotes the kinematic viscosity coefficient.

For the time integration method, the prediction-correction method was used to iterate in this study.

The prediction step:
$$\begin{cases}
\rho_I^{n+\frac{1}{2}} = \rho_I^n + \dfrac{\Delta t}{2}\left(\dfrac{d\rho_I}{dt}\right)^n \\
v_I^{n+\frac{1}{2}} = v_I^n + \dfrac{\Delta t}{2}\left(\dfrac{dv_I}{dt}\right)^n \\
x_I^{n+\frac{1}{2}} = x_I^n + \dfrac{\Delta t}{2} v_I^{n+\frac{1}{2}}
\end{cases} \tag{28}$$

The correction step:
$$\begin{cases}
\rho_I^{n+1} = \rho_I^n + \Delta t\left(\dfrac{d\rho_I}{dt}\right)^{n+\frac{1}{2}} \\
v_I^{n+1} = v_I^n + \Delta t\left(\dfrac{dv_I}{dt}\right)^{n+\frac{1}{2}} \\
x_I^{n+1} = x_I^n + \Delta t v_I^{n+1}
\end{cases} \tag{29}$$

where $\Delta t$ is the time step size, and in the prediction step, the physical information of the iterative half time step is obtained. Then, in the calibration step, the density, velocity and temperature change rate of the particles are calculated according to the position and physical quantity of the particles in the estimated step, and then the density, velocity, temperature and particle position are advanced by one time step from the reference step.

## 3. Numerical examples

In order to verify the feasibility of the proposed method, we have done several typical cases to verify the feasibility of ULPH heat transfer, including Couette flow and Poiseuille flow. The models used are adiabatic models in this work, and we do not consider the problem of thermal radiation and energy loss.

### 3.1 Accuracy and convergence

Firstly, we verify the accuracy and convergence of the second-order ULPH theory. The basic mathematical function $f(x,y) = e^{-(x^2+y^2)}$ is selected as the model for convergence verification, and the computational domain is $(x, y) \in [-1,1] \times [-1,1]$. The domain is discretized into particles with the uniform and non-uniform distributions as shown in Fig. 2.

It is worth mentioning that we also consider the boundaries of the functional model and add three additional layers of boundary particles to the computational domain (Fig. 2) to solve the problem of missing neighbors of particles near the boundary. The role of the boundary particle is only used to solve for the shape tensor M of the particle in the computational domain close to the boundary.

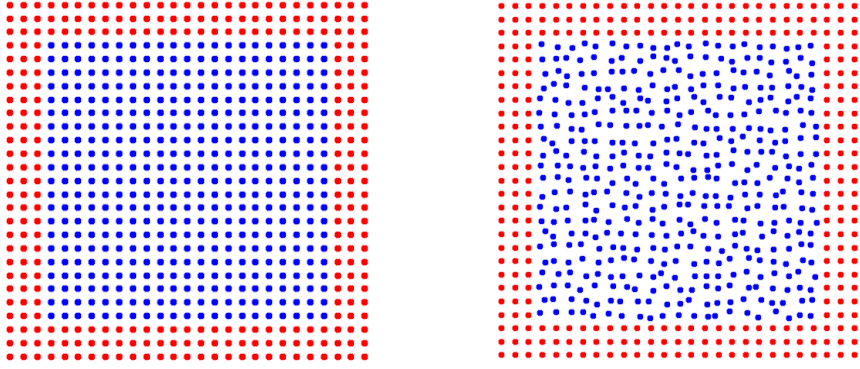

Fig. 2 Uniform and non-uniform particle distribution (red for boundary particles)

The convergence speed of a given function is measured by calculating the root mean square error (RMSE), which is defined as:

$$\text{RMSE}(f) = \left[ \sum_{J=1}^{N} \frac{1}{N} \left( f_J^{\text{predicted}} - f_J^{\text{exact}} \right)^2 \right]^{0.5}$$

where $N$ is the total number of particles in computational domain, $f^{\text{predicted}}$ and $f^{\text{exact}}$ represent the numerical and exact solutions, respectively. The entire domain is discretized into 1681 particles with uniform and non-uniform distributions. Table 1 shows the errors of the first and second derivatives of the basis of the quadratic polynomial, and compares them with the results of Yan[22] without considering the boundaries, where the action of boundary particles can indeed significantly reduce the errors. Non-local differential operators have high accuracy in both uniform and inhomogeneous particle distributions.

Table 1 The RMSE of the derivatives with uniform and non-uniform particle distributions. (The resolution is 41*41=1681)

| Derivatives | Uniform | | Non-uniform | |
| --- | --- | --- | --- | --- |
| | Consider boundaries | No boundaries | Consider boundaries | No boundaries |
| $f_x$ | 0.00181125 | 0.00250 | 0.00182584 | 0.00258 |
| $f_y$ | 0.00181125 | 0.00250 | 0.00182314 | 0.00257 |
| $f_{xx}$ | 0.00321437 | 0.01771 | 0.00378575 | 0.01837 |
| $f_{xy}$ | 0.00241138 | 0.00593 | 0.00342449 | 0.00730 |
| $f_{yy}$ | 0.00321437 | 0.01771 | 0.00379594 | 0.01846 |

In order to verify the convergence speed of the second-order non-local differential operator of ULPH, we use the distribution of uniform and non-uniform particles with resolutions ranging from 121 to 1002001 in the computational domain. The convergence speed of the first derivative $f_x$ and $f_y$ is shown in Fig. 3(a). It can be seen that the root mean square error decreases exponentially with the increase of the number of particles, regardless of whether it is uniformly distributed or non-uniformly distributed, and the error of uniform distribution is slightly lower than that of non-uniform distribution. The convergence of the second derivative is shown in Fig. 3(b). It can be seen that as the number of particles increases, the accuracy of the second derivative also increases. The particle distribution has little effect on the convergence speed of the second-order non-local differential operator of ULPH, and the error is maintained at a low level for both the first and second derivatives. These results demonstrate the accuracy and effectiveness of the second-order non-local differential operator in solving the derivative of the function.

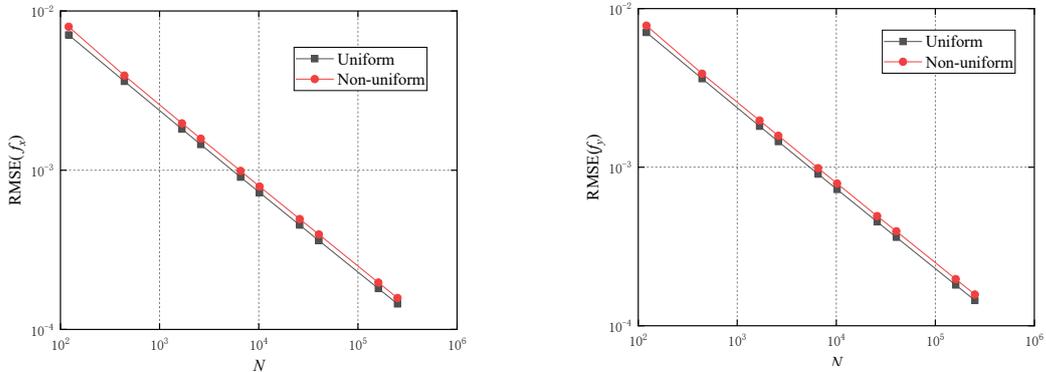

(a) RMSE of the first derivative of uniform and non-uniform distribution

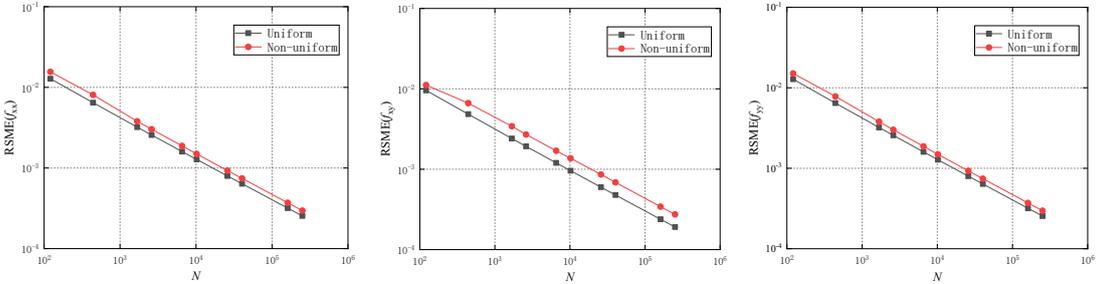

(b) RMSE of the second derivative of uniform and non-uniform distribution

Fig. 3 Convergence rates different particle numbers with uniform distribution and non-uniform distribution.

## 3.2 Transient heat conduction

In this section, the ULPH heat transfer equation is used to simulate the heat conduction of hydrostatic water in a 0.1m × 0.1m square cavity. The square chamber is filled with 10 °C water, and the left wall is set to a constant temperature of 100 °C, the rest of the boundaries are kept at a constant temperature of 10 °C (Fig. 4). The fluid density is $1000\ kg \cdot m^{-3}$, the specific heat capacity is $4182\ J \cdot kg^{-1} \cdot K$, and the heat transfer coefficient is $0.6\ W \cdot m^{-1} \cdot K$.

Comparing the calculated results with the results of the FVM and SPH, in SPH and ULPH simulations, the particle spacing is $dx$=0.0005m, and the mesh size selected in FVM is also 0.0005m to ensure the same discretization conditions. Fig. 5 shows a comparison of the heat transfer results of the three methods at different times (200s, 500s, 1000s) and the results indicate a high level of concordance. Fig. 6 shows the temperature distribution on the central axis at the time corresponding to Fig. 5, the outcomes exhibit robust agreement, indicating that ULPH can be used to accurately model heat transfer processes in liquid media.

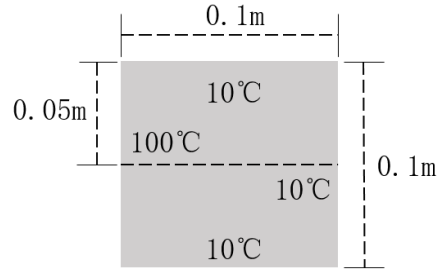

Fig. 4 Hydrostatic heat conduction model in a square cavity

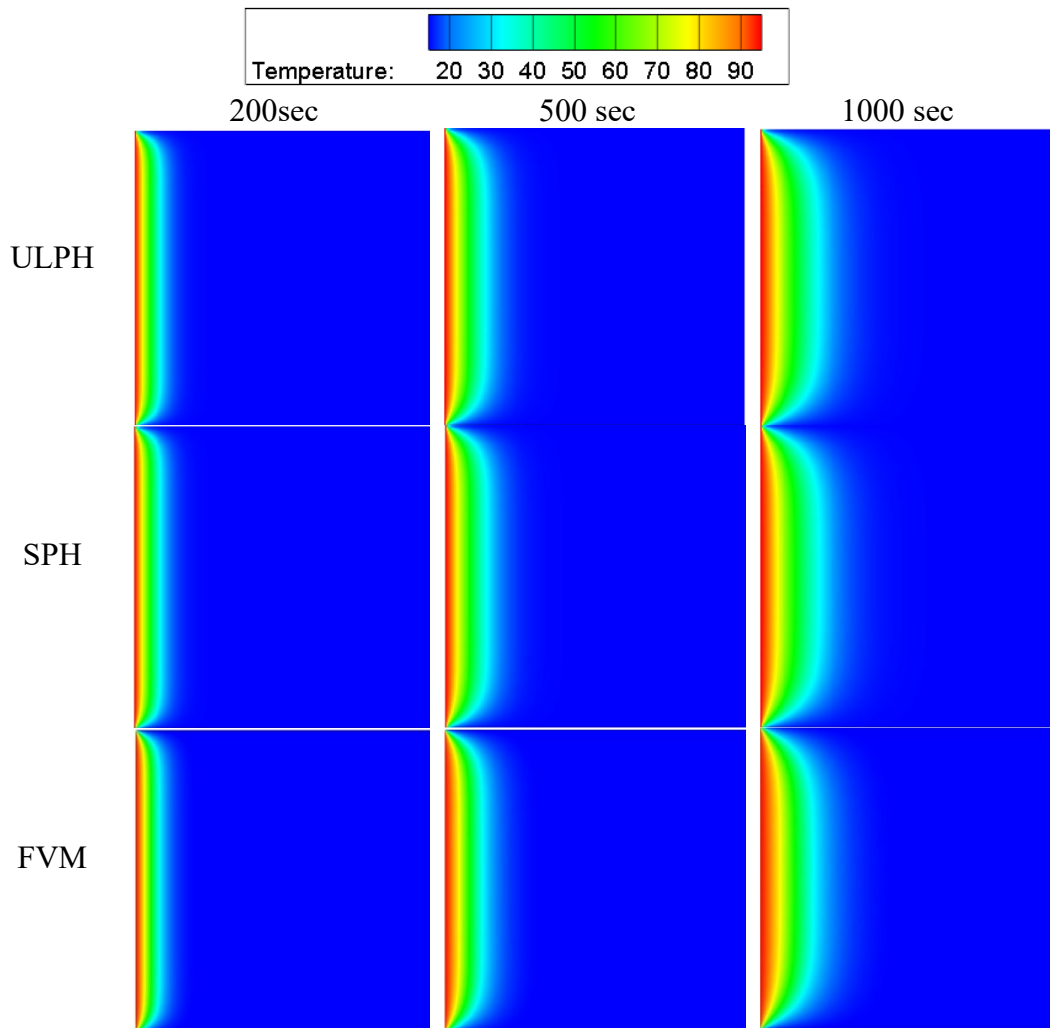

Fig. 5 Comparison of hydrostatic heat conduction results in square cavities at different times (ULPH、SPH、FVM)

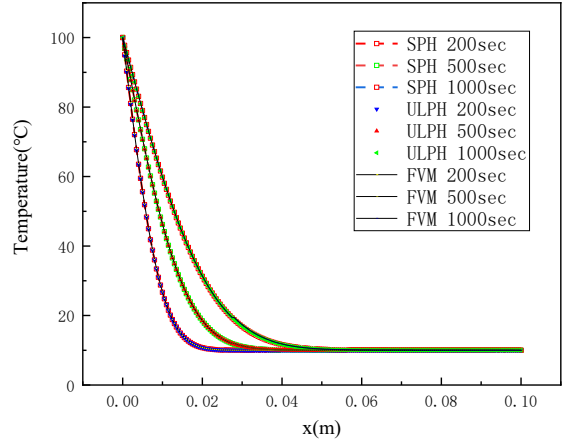

Fig. 6 Temperature distribution on the axis of y=0.

## 3.3 Transient heat convection

Convective heat transfer problems are prevalent in industrial and engineering applications, such as chip cooling, nuclear reactor cooling, liquid heating, air cooling and heat dissipation, and so on. In this section, we use the Couette flow and the Poiseuille flow problem to verify the convective heat transfer capacity of the proposed method. Both cases are simulated under laminar conditions。

### 3.3.1 Transient Couette flow

In this section, we consider the Couette flow problem involving heat transfer to verify the feasibility of the proposed method in the transient heat transfer problem. The model (Fig. 7) is a two-dimensional rectangular channel of 0.1 m × 0.2 m with a particle spacing of 0.005m. The fluid density is 1.0 $kg \cdot m^{-3}$, the kinematic viscosity $\mu = 0.001 m^2 s^{-1}$, the specific heat capacity is 1000, and the heat transfer coefficient is $0.1 W \cdot m^{-1} \cdot K$.

The boundary between the upper and lower solid walls is set as a non-slip boundary, and the inflow and outflow are treated as periodic boundaries. The top (T=100°C) and bottom (T=10°C) boundaries are set to isothermal boundaries, and the fluid particles have an initial temperature of 10°C and are at rest. The constant velocity $u_0$=0.05m/s of the upper plate drives the fluid motion, and the simulation is performed until t=30 s to ensure that the steady state is reached.

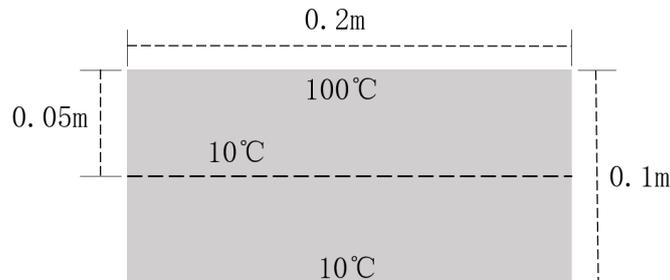

Fig. 7 Couette flow model

When the computation reaches steady state, the velocity and temperature should be consistent with the analytical solution and have a linear distribution:

$$u(y) = 0.5(y + 0.05)$$
$$T(y) = 450(y + 0.05) + 10$$

where $u(y)$ is the velocity distribution function and $T(y)$ is the temperature distribution function.

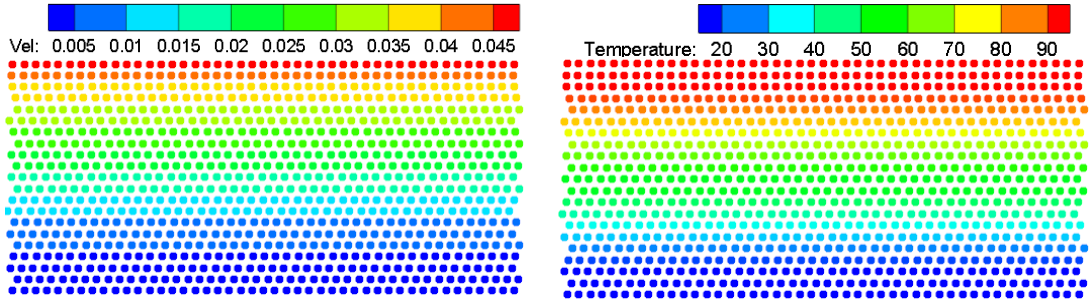

Fig. 8 Schematic diagram of the steady-state velocity and temperature results of the Couette flow

Fig. 8 shows the temperature and velocity distribution after reaching steady state. It can be seen that the temperature and velocity of the fluid particles are linearly distributed along the vertical direction of the plate. Fig. 9 shows the velocity and temperature distribution at x=0.1m, which are in perfect agreement with the analytical solution.

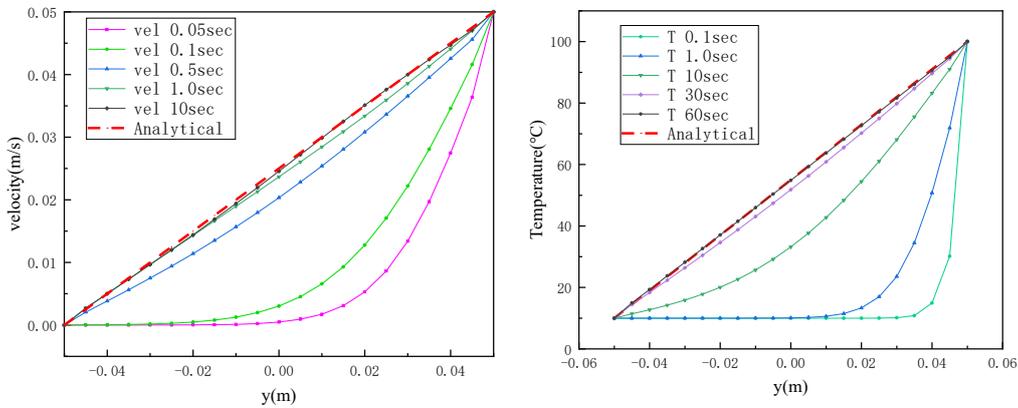

Fig. 9 Velocity and temperature distribution curves at x = 0.1m

### 3.3.2 Transient Poiseuille flow

In this section, we consider the Poiseuille flow problem involving heat transfer to verify the feasibility of the proposed method in transient heat transfer problems. The model we used is a 2D rectangular channel (Fig. 10) of 0.1m × 1.0 m with a particle spacing of 0.005 m. The fluid density is 100.0 $kg \cdot m^{-3}$, the kinematic viscosity $\mu = 0.001 m^2 s^{-1}$, the specific heat capacity is 1000 $J \cdot kg^{-1} \cdot K$ and the heat transfer coefficient is 10 $W \cdot m^{-1} \cdot K$.

The acceleration loading in the horizontal direction of 0.1 $ms^{-2}$ is adopted. The solid wall boundaries are set as non-slip boundaries, and the inflow and outflow are treated as periodic boundaries. The top and bottom boundaries are set to isothermal boundaries (T=100°C). The outflow boundary is considered adiabatic and the temperature of the inflow boundary is fixed at T = 10 °C. The simulation was performed until t=30 s to ensure that a steady state was reached.

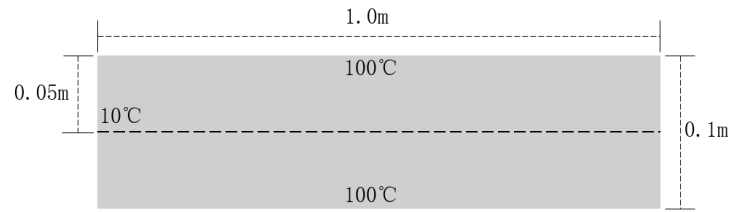

Fig. 10 Poiseuille flow model

When the computation reaches steady state, the velocity distribution should be consistent with the analytic solution, showing a parabolic distribution:

$$u = u_m \left[1 - \left(\frac{r}{R}\right)^2\right]$$

where $u$ is the velocity distribution function.

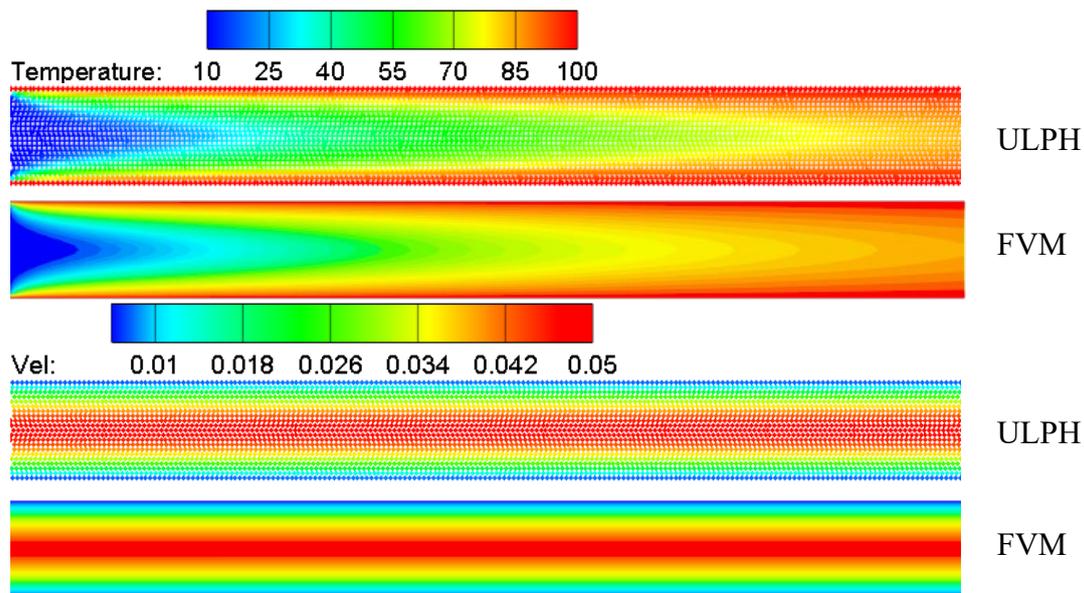

Fig. 11 Steady-state Temperature and velocity distributions of the transient Poiseuille flow

Fig. 11 shows the temperature and velocity distribution after reaching steady state. Under the flow conditions of the expected full development, the uniform profile is obvious. Due to the presence of the heating wall, the fluid temperature rises monotonically along the flow direction, and a significant convective heat transfer effect can be seen. Superficially, the thickness of the thermal boundary layer also grows along the direction of flow.

To verify the accuracy of the simulation results, we compared the temperature distributions of different cross-sections from x=0.1m to x=0.9m at steady state with the results of the FVM, as shown in Fig. 12. As can be seen from Fig. 12, the temperature gradient decreases as fluid particles flow downstream due to the thickening of the thermal boundary layer. The results show that the transient heat transfer Poiseuille flow problem solved by the second-order ULPH is in full agreement with the results of the finite volume method. For the problem of convective heat transfer, ULPH has applicability, compared with SPH, ULPH does not need to be varied on the formula, and can meet the requirements of arbitrary order accuracy.

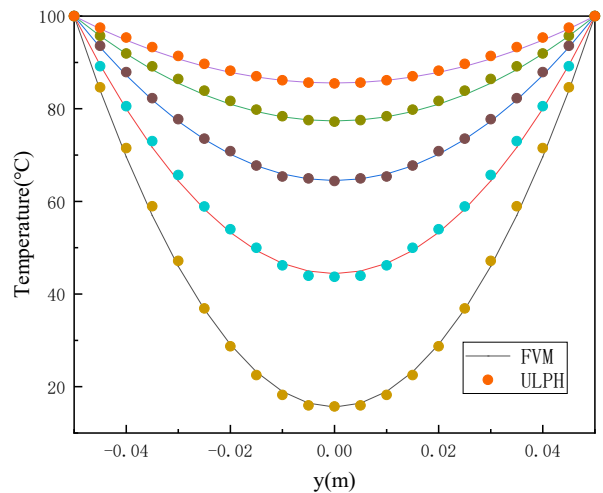

Fig. 12 Steady-state temperature distributions at x=0.1m, 0.3m, 0.5m, 0.7m and 0.9m.

## 4. Conclusion

In this paper, the ULPH heat transfer equation is derived, which is used to model heat conduction and heat convection under laminar flow conditions. The simulated cases include the hydrostatic heat transfer in square cavity, Couette flow and Poiseuille flow. The same cases were also simulated using FVM to benchmark the solution.

The results show the accuracy of the proposed method. It is shown that ULPH has good potential in solving the problem of convective heat transfer. ULPH performs exceptionally well, exhibiting excellence in both accuracy and convergence, and is minimally influenced by the manner in which particles are distributed. As a starting point for the ULPH study of heat transfer, thermal radiation can be considered in the future, also involving fluid-solid conjugate heat transfer. The aim of this work is to further expand and apply the ULPH theory. As a new meshless method, there is more potential value in ULPH.

The ULPH heat transfer model used in this work requires further investigation of turbulence and will be extended to more realistic engineering problems in the future.